\documentclass[12pt]{article}
\usepackage{amssymb, graphicx, hyperref}
\begin{document}
\title{\bf\Large Interactions between Dielectric Branes
\vspace{28pt}}
\author{\vspace{8pt}\normalsize 
Yi-hong Gao$^{(a)}$\footnote{e-mail: {\tt gaoyh@itp.ac.cn}}  \
and Zhong-xia Yang$^{(b)}$\footnote{e-mail: {\tt zxyang@pine.njnu.edu.cn}}
\\ 
$^{(a)}${\it\small Institute of Theoretical Physics, Beijing 100080, 
            P.~R.~China}\\
 $^{(b)}${\it\small Department of Physics, Nanjing Normal University, 
            P.~R.~China}}
\date{}
\maketitle
\baselineskip .64cm
\vskip 1.6cm
\centerline{\bf Abstract}
\vskip .3cm
Quantum corrections to the energy of D0-branes in a constant RR 4-form background are studied. Using the Born-Oppenheimer approximation, we compute the long-range interaction between two spherical D2-D0 bound states. We extend this calculation to the case where some mass terms are added. For the special value of masses at which the classical energy vanishes, we find complete cancellations of two boson-fermion pairs in the quantum mechanical expression of the zero-point energy, suggesting possible restoration of (partial) supersymmetries. We also briefly discuss the interaction between a dielectric 2-brane and a single D0-brane.
\renewcommand{\theequation}{\thesection.\arabic{equation}}
\csname @addtoreset\endcsname{equation}{section}
\newcommand{\mbb}[1]{\mathbb{#1}}
\vskip 3.5cm
\newpage
\voffset -.2in
\section{Introduction}
D-branes in the presence of higher rank RR background fields \cite{TR} \cite{Myers} may look somewhat unusual. If one couples $N$ D0-branes to a constant 4-form field strength $F$, for example, then these D particles will blow up into a dielectric 2-brane, namely a D0-D2 bound state, whose geometrical configuration is characterized by some fuzzy 2-sphere of radius $\sim NF$ \cite{Myers}. Such a phenomenon (and its higher dimensional analog) is known as Myers' dielectric effect. 

The Myers effect has raised a number of interesting issues \cite{MST}--\cite{GR}. A similar phenomenon was found \cite{MST} for a graviton propagating in the spherical part of the product space $AdS\times S$ with a nonvanishing angular momentum. The graviton will expand into a spherical brane, or ``giant graviton'', whose size grows as the angular momentum is increased. Since this size is bounded by the size of the surrounding sphere, the angular momentum must have a maximal value. This therefore gives a clear physical illustration of how the stringy exclusion principle \cite{MS} works. As elaborated in \cite{Miao}, the same effect is also consistent with the spacetime uncertainty principle \cite{JYMY}. The relation between giant gravitons and the magnetic analog of Myers' mechanism has been explored in \cite{DTV}. In addition, supergravity solutions of dielectric branes can be constructed explicitly in the $AdS$ background, which provide a remarkable dual description of a certain confining gauge theory in four dimensions \cite{PS}. 

In this paper we consider two-body interactions between the D0-D2 bound states. The one-body potential of such objects can be found in \cite{Myers}, which shows that the ground state energy has a negative value, so SUSY is completely lost even at the tree level. It also indicates that the configurations of two or more dielectric branes are classically unstable, --- they tend to form a single, larger fuzzy $S^{2}$. In other words, given the number $N$ of D0-branes coupled to the RR field $F$, the classical solution in the $N$-dimensional irreducible representation of $SU(2)$ has the lowest energy, lower than those in any reducible ones. However, if the potential is deformed by some mass terms, then it is possible to choose a suitable mass parameter so that the ground state energy vanishes for all $N$-dimensional representations. This corresponds, in the AdS/CFT context, to the interesting case where ${\cal N}=1^{*}$  SUSY preserving mass terms are added to the ${\cal N}=4$ theory \cite{PS}. In this situation all the static solutions are degenerate and we can get classically stable configurations of multi dielectric branes. Interactions between these branes may thus arise as purely a quantum mechanical effect.

Our quantum mechanical calculations will follow \cite{AB} closely. We start with the matrix model expression of the energy, perturbing it around the classical configuration of two dielectric branes separated by a distance $r$, and keep terms up to the second order of fluctuations. According to the Born-Oppenheimer approximation, we treat off-diagonal blocks in the perturbed matrices as fast modes, and quantize them under a field theory limit. Quantum corrections to the zero-point energy can then be read off from the mass matrix of these off-diagonal modes. Similar methods have been used in the matrix model computation of membrane interactions \cite{AB} \cite{LM} \cite{TR97} (see \cite{Taylor} for a recent review).

Before actual computations let us give a very rough estimate of how the potential $V(r)$ between two D0-D2 bound states should behave. Each such dielectric 2-brane could be seen as a cluster of $N$ D0-branes, with the cluster-cluster interaction of the form \cite{DKPS}
\begin{equation}
V(r,v)\sim N^{2}\frac{v^{4}}{r^{7}},
\label{cluster}
\end{equation}
where $v$ is the relative velocity. To determine $v$, assume further that the two clusters form a bound state, so that the Viral theorem applies to the matrix model expression of the total energy. The kinetic energy may be estimated as $NT_{0}v^{2}/2$, with $T_{0}=g_{s}^{-1}l_{s}^{-1}$ being the tension of D0-branes, while the RR background $F$ will contribute to the matrix model potential by a term $NT_{0}Fr^{3}/(2\pi l_{s}^{2})$. These two terms should roughly have the same value and hence we get a relation $v^{2}\sim Fr^{3}$ in string units. Substituting this into (\ref{cluster}) yields
\begin{equation}
V(r)\sim N^{2}F^{2}/r.
\label{1/r}
\end{equation}

As we shall see, the $1/r$-behavior of the interactions agrees with our detailed quantum mechanical calculations, at least for both $r$ and $N$ being large\footnote{Actually, even the factor $F^{2}$ in (\ref{1/r}) is correct, which can be reproduced by our more serious computations.}. Such an $r$-dependence could be expected from a rather general argument presented in Section 3.3 of \cite{AB}. In fact, the zero-point energy is the sum of bosonic frequencies minus the sum of fermionic frequencies
\begin{equation}
V(r)\sim\sum_{\bf k}\left[\sum_{i=1}^{n_{b}}a_{i}({\bf k})\sqrt{r^{2}+b_{i}({\bf k})}-\sum_{i=1}^{n_{f}}c_{i}({\bf k})\sqrt{r^{2}+d_{i}({\bf k})}\right],
\label{generic}
\end{equation}
where $n_{b}$, $n_{f}$ denote the numbers of bosons and fermions, respectively (in our system both $n_{b}$ and $n_{f}$ turn out to be 8), and $\bf k$ labels the quantized modes. Now for a generic theory with equal numbers of bosons and fermions, the long-range potential depends linearly on $r$. However, if this theory comes from a supersymmetric action with some additional SUSY-violation terms, such as the coupling of RR 4-form $F$ to D0-branes, we expect $a_{i}({\bf k})=c_{i}({\bf k})$ and the potential will decay no slower than $1/r$ due to cancellations between the bosonic and fermionic leading terms in (\ref{generic}). More cancellations could occur if the underlying system is actually supersymmetric. For instance, the zero-point energy of BPS states in a theory with maximal SUSY's vanishes because apart from $a_{i}({\bf k})=c_{i}({\bf k})$, we also have $b_{i}({\bf k})=d_{i}({\bf k})$ for all modes. Even for some configurations with broken SUSY, such as the membrane/anti-membrane system studied in \cite{AB}, there are certain degrees of matching between $b_{i}({\bf k})$ and $d_{i}({\bf k})$, leading to a behavior $V(r)\sim 1/r^{l}$, $l > 1$. For dielectric branes, of course, the amount of such cancellations should be much less than those considered in \cite{AB}. That (\ref{1/r}) behaves as $1/r$ corresponds to the case of least cancellations.

At this point, one may ask what happens if the action of dielectric branes is perturbed by some mass terms. When the mass parameter is chosen such that the minimum of the classical potential vanishes, we expect that a kind of supersymmetry, similar to ${\cal N}=1^{*}$ SUSY in 4 dimensions \cite{PS}, will be restored and some additional cancellations will occur in (\ref{generic}). We shall actually find that two of the eight bosonic objects $b_{i}({\bf k})$, say $b_{1}({\bf k})$ and $b_{2}({\bf k})$, are precisely equal to the corresponding fermionic ones for all quanta $\bf k$, namely $b_{1}({\bf k})\equiv d_{1}({\bf k})$, $b_{2}({\bf k})\equiv d_{2}({\bf k})$. This ensures that all the relevant bosonic and fermionic modes cancel exactly with each other. The resulting interaction is thus weaker by some numerical factor, but it should still go like $1/r$ since there are no further cancellations between the remaining six $b_{i}({\bf k})$'s and $d_{i}({\bf k})$'s.

It would be interesting to understand the $1/r$ behavior from the point of view of M-theory or its supergravity approximations. But as far as we can say at present, formulating M(atrix) theory in background fields is a rather subtle problem \cite{DOS}, and its success relies essentially on the existence of a decoupling limit \cite{BFSS}, which may no longer exist if a nontrivial background is introduced. It is far from clear yet whether a given background is consistent beyond low-energy approximations. This problem is closely related to background independence and covariantizing the Matrix Dynamics.

The present paper is organized as follows. In Section~\ref{s2} we investigate quantum corrections to the long-range potential between two dielectric 2-branes, focusing on the distance dependent part of the interactions. In Section~\ref{s3}, this computation is extended to the case where some mass terms are added. Section~\ref{s4} then provides a brief discussion of the interaction between a dielectric 2-brane and a single D0-brane. Finally, Section~\ref{s5} summarizes our conclusions.
\section{Quantum Mechanics of Dielectric 2-Branes}\label{s2}
We begin with setting up the classical configuration of two dielectric 2-branes separated by a distance $r$, and then turn to its quantum fluctuations. In an RR 4-form background $F^{(4)}$, the static configuration of $N$ D0-branes has the potential energy \cite{Myers}
\begin{equation}
V = -\frac{T_{0}}{4\lambda^{2}}\sum_{a\neq b}{\rm Tr}[X^{a},X^{b}]^{2}
-\frac{iT_{0}}{3\lambda}{\rm Tr}(X^{i}X^{j}X^{k})F^{(4)}_{0ijk}-\frac{T_{0}}{\lambda}{\rm Tr}(\theta^{T}\gamma_{a}[X^{a},\theta])
\label{energy1}
\end{equation}
where $X^{a}$ are $N\times N$ hermitian matrices with fermionic partners $\theta$, and $\lambda=2\pi l_{s}^{2}$. Since the term proportional to $-{\rm Tr}[X^{a},X^{b}]^{2}$ is non-negative, for $F^{(4)}=0$ the potential is minimized by mutually commuting matrices. In this case $V$ vanishes when its minimum is arrived. If a nonvanishing $F^{(4)}$-term is present, however, the equations of motion for the static configuration will become
\begin{equation}
[[X^{i},X^{j}],X^{j}]-\frac{i\lambda}{2}F^{(4)}_{0ijk}[X^{j},X^{k}]=0,
\label{EOM1}
\end{equation}
which allow, in addition to the mutually commuting solution mentioned above, some other more interesting solutions giving $V$ a negative value.

Consider the case when these D0-branes live in a constant RR background:
\begin{equation}
F^{(4)}_{0ijk}=\left\{\begin{array}{ll}\displaystyle{
-\frac{2}{\lambda}f\epsilon_{ijk}},\quad &\hbox{\rm if }1\leq i,j,k\leq 3\\
0, & {\rm otherwise}.
\end{array}
\right.
\label{R-R}
\end{equation}
Here the constant $f$ has dimension of length. Let $J_{i}$ be a basis of $SU(2)$ generators, satisfying the commutation relations
$
[J_{i},J_{j}]=i\epsilon_{ijk}J_{k}
$.
One finds that $X^{i}=fJ_{i}$ solves the equations (\ref{EOM1}).
We can restrict ourselves to an $N$-dimensional representation of $J_{i}$ and take the classical configuration to be
\begin{equation}
X^{i}=f J_{i},\quad i=1,2,3;\quad\quad X^{\mu}=0,\quad\mu\geq 4.
\label{ansatz}
\end{equation}
Suppose that this representation is irreducible, namely, it is labelled by a spin $j=0,\frac{1}{2},1,\cdots$, with
$2j+1=N$ (or $j=\frac{N-1}{2}$). Then, for $X^{i}=fJ_{i}$, the potential (\ref{energy1}) takes its minimal value
\begin{equation}
V=-\frac{T_{0}f^{4}}{6\lambda^{2}}NC^{(j)}_{2}=-\frac{T_{0}f^{4}}{24\lambda^{2}}N(N^{2}-1),
\label{min}
\end{equation}
where
\begin{equation}
C_{2}^{(j)}=j(j+1)=\frac{N^{2}-1}{4}
\label{casimir}
\end{equation}
is the second order Casimir $C_{2}=\sum_{i=1}^{3}(J_{i})^{2}$ in that irreducible representation.

Thus, given $N$, the classical configuration (\ref{ansatz}) for a dielectric 
brane constitutes a fuzzy sphere \cite{Myers}. Note that if $(X^{1},X^{2},X^{3})$ is a solution to the classical equations 
of motion, then the ``anti'' dielectric brane $(-X^{1},-X^{2},-X^{3})$ does 
not solve the same equations, unless in the degenerate 
case $f=0$. In other words, the system we are considering does not allow 
anti dielectric brane solutions classically. Physically this should be quite clear: unlike the usual D2-branes that must carry RR charges, dielectric 2-branes are neutral, carrying only dipole (or multipole) moments \cite{Myers}.
\subsection{Parallel Dielectric Branes}
Now consider a system consisting of two dielectric branes, parallel to 
each other and with some distance $r$ apart along the $4^{th}$ direction. Classically, 
this state is described by the configuration
\begin{equation}
X^{i}=\pmatrix{fJ^{(1)}_{i} & 0 \cr
	0 & fJ^{(2)}_{i}}\quad (1\leq i,j,k\leq 3),\quad\quad
	X^{4}=\pmatrix{0 & 0 \cr
	0 & r},\quad\quad X^{\mu}=0 \quad (\mu\geq 5)
\label{config}
\end{equation}
here $J^{(1)}_{i}$ and $J^{(2)}_{i}$ are two independent sets of the $SU(2)$ 
Lie algebra generators, each belonging to a copy of the irreducible representation of spin $j=(N-1)/2$. Clearly, such $X^{i}$ are all block diagonal $2N\times 2N$ matrices, valued in the reducible representation $j\oplus j$ of $SU(2)$.

At the classical level, the energy of this configuration is computed by a sum 
of two copies of (\ref{min}), and the result reads
\begin{equation}
V_{2N}=-\frac{T_{0}f^{4}}{12\lambda^{2}}N(N^{2}-1).
\label{2Npotential-reducible}
\end{equation}
Although this energy takes a negative value, it is still higher than the 
energy of a dielectric brane described by some matrices $X^{i}$ belonging to 
the $2N$-dimensional irreducible representation of 
$SU(2)$ \cite{Myers}. Actually, the latter energy is determined by 
substituting $N\rightarrow 2N$ into (\ref{min}), which gives
\begin{equation}
V_{2N}^{\rm irr}=-\frac{T_{0}f^{4}}{12\lambda^{2}}N(4N^{2}-1).
\label{2Npotential-irr}
\end{equation}
Hence the difference between (\ref{2Npotential-reducible}) 
and (\ref{2Npotential-irr}) is
\begin{equation}
V_{2N}-V_{2N}^{\rm irr}=\frac{T_{0}f^{4}}{4\lambda^{2}}N^{3} > 0.
\end{equation}
It follows that classically (\ref{config}) is a meta-stable configuration, so that two parallel dielectric branes of ``size'' $N$ tend to form a single, 
larger dielectric brane of ``size'' $2N$.
\subsection{Quantum Corrections}
We now look at quantum corrections to the potential of two parallel dielectric 2-branes. We will expand the energy (\ref{energy1}) around its classical configuration (\ref{config}), keeping terms up to the second order of fluctuations. Such fluctuations are not necessarily static, and there are  contributions also from terms involving the momentum $\Pi_{i}\sim T_{0}\dot{X}^{i}$. The zero-point energy is then determined by the frequencies of all bosonic and fermionic oscillators. Our main task here is to find these frequencies, which are proportional to the eigenvalues of certain mass matrices.

Let us introduce dimensionless quantities $Y^{a}$, $\theta'$ and $r'$, via
\begin{equation}
X^{a}=fY^{a},\quad r=fr', \quad \theta=T_{0}^{-1/2}\theta'.
\label{noM}
\end{equation}
As in \cite{AB}, we only consider off-diagonal quantum fluctuations.
Thus, in terms of the dimensionless variables $Y^{a}=(Y^{i},Y^{4},Y^{\mu})$ (with $1\leq i\leq 3$ and $5\leq \mu\leq 9$), we take:
\begin{equation}
Y^{i}=\pmatrix{J^{(1)}_{i} & A_{i}\cr
                         A_{i}^{\dag} & J^{(2)}_{i}},\quad
Y^{4}=\pmatrix{0 & A_{4}\cr
                         A_{4}^{\dag} & r'},\quad
Y^{\mu}=\pmatrix{0 & A_{\mu}\cr
                         A_{\mu}^{\dag} & 0}.
\label{bFluc}
\end{equation}
Similarly, quantum fluctuations of fermionic degrees of freedom are taken to be
\begin{equation}
\theta'=\pmatrix{0 & \psi\cr
                0 & 0}, \quad \theta'^{T}=\pmatrix{0 & 0\cr
                \psi^{T} & 0}
\label{fFluc}
\end{equation}
where the upper-right block $\psi$ constitutes a fermionic $N\times N$ matrix.
\subsubsection{Quadratic Terms of the Off-diagonal Modes}
From the ansatz above, one can compute various terms in the potential (\ref{energy1}) up to the second order of $A_{a}$ and $\psi$. For terms with bosonic degrees of freedom only, we obtain:
\begin{eqnarray}
\sum_{1\leq i,j\leq 3}{\rm Tr}[Y^{i},Y^{j}]^{2}&=&{\rm Tr}\left([J^{(1)}_{i},J^{(1)}_{j}]^{2}+[J^{(2)}_{i},J^{(2)}_{j}]^{2}\right)\nonumber\\
&-&4{\rm Tr}\left(A_{i}^{\dag}J^{(1)}_{j}J^{(1)}_{j}A_{i}+ A_{i}J^{(2)}_{j}J^{(2)}_{j}A_{i}^{\dag}-2J^{(1)}_{i}A_{j}J^{(2)}_{i}
A_{j}^{\dag}\right.\nonumber\\
&+&J^{(1)}_{i}A_{j}J^{(2)}_{j}A_{i}^{\dag}+J^{(1)}_{i}A_{i}J^{(2)}_{j}A_{j}^{\dag}
+J^{(1)}_{i}A_{i}A_{j}^{\dag}J^{(1)}_{j}\nonumber\\
&+&\left. J^{(2)}_{i}A_{i}^{\dag}A_{j}J^{(2)}_{j}-2J^{(1)}_{i}A_{j}
A_{i}^{\dag}J^{(1)}_{j}-2J^{(2)}_{i}A_{j}^{\dag}A_{i}J^{(2)}_{j}\right),
\label{Bmass1}
\end{eqnarray}
\begin{eqnarray}
\sum_{i=1}^{3}{\rm Tr}[Y^{i},Y^{4}]^{2} &=& -2{\rm Tr}\left(J^{(1)}_{i}A_{4}A_{4}^{\dag}
J^{(1)}_{i}+A_{4}J^{(2)}_{i}J^{(2)}_{i}A_{4}^{\dag}
-2J^{(1)}_{i}A_{4}J^{(2)}_{i}A_{4}^{\dag}
\right.\nonumber\\
&+&\left. r'J^{(1)}_{i}A_{4}A_{i}^{\dag}
+r'J^{(1)}_{i}A_{i}A_{4}^{\dag}-r'A_{i}J^{(2)}_{i}A_{4}^{\dag} 
- r'A_{4}J^{(2)}_{i}A_{i}^{\dag} \right.\nonumber\\
&+& \left.r'^{2}A_{i}A^{\dag}_{i}\right),
\label{Bmass2}
\end{eqnarray}
\begin{equation}
\sum_{i\leq 3,\;\mu\geq 5}{\rm Tr}[Y^{i},Y^{\mu}]^{2}=-2{\rm Tr}
\left(J^{(1)}_{i}A_{\mu}A_{\mu}^{\dag}J^{(1)}_{i}
+A_{\mu}J^{(2)}_{i}J^{(2)}_{i}A_{\mu}^{\dag}
-2J^{(1)}_{i}A_{\mu}J^{(2)}_{i}A_{\mu}^{\dag}\right),
\label{Bmass3}
\end{equation}
\begin{equation}
\sum_{\mu\geq 5}{\rm Tr}[Y^{4},Y^{\mu}]^{2}=-2{\rm Tr}\left(r'^{2}A_{\mu}A^{\dag}_{\mu}\right),
\label{Bmass4}
\end{equation}
and
\begin{eqnarray}
{\rm Tr}(Y^{i}Y^{j}Y^{k})\epsilon_{ijk} &=&{\rm Tr}\left(J_{i}^{(1)}J_{j}^{(1)}J_{k}^{(1)}+J_{i}^{(2)}J_{j}^{(2)}J_{k}^{(2)}\right)\epsilon_{ijk}\nonumber\\
&+&3{\rm Tr}\left(J^{(1)}_{i}A_{j}A^{\dag}_{k}-J^{(2)}_{i}A^{\dag}_{j}A_{k}\right)\epsilon_{ijk}.
\label{Bmass5}
\end{eqnarray}
Summing over (\ref{Bmass1})-(\ref{Bmass5}) with appropriate constant coefficients yields the bosonic part of the potential (\ref{energy1}), up to  quadratic terms of the off-diagonal modes. 

Note that the leading terms in (\ref{Bmass1}) and (\ref{Bmass5}) are independent of $A_{a}$, whose values can be evaluated via the following simple formulae
\begin{equation}
{\rm Tr}[J^{(1)}_{i},J^{(1)}_{j}]^{2}={\rm Tr}[J^{(2)}_{i},J^{(2)}_{j}]^{2}=
-2NC_{2}^{(j)}=-\frac{N(N^{2}-1)}{2},
\end{equation}
\begin{equation}
{\rm Tr}\left(J^{(1)}_{i}J^{(1)}_{j}J^{(1)}_{k}\right)\epsilon_{ijk}
={\rm Tr}\left(J^{(2)}_{i}J^{(2)}_{j}J^{(2)}_{k}\right)\epsilon_{ijk}
=iNC_{2}^{(2)}=\frac{i}{4}N(N^{2}-1).
\end{equation}
Such terms determine the classical minimum (\ref{2Npotential-reducible}) of the potential. Quantum mechanically, the zero-point energy of bosonic off-diagonal modes arises from the mass matrix of $A_{a}$, which comes out from the quadratic terms in (\ref{Bmass1})-(\ref{Bmass5}). 

On the other hand, for terms in the potential that involve fermionic degrees of freedom, one finds
\begin{equation}
{\rm Tr}(\theta'^{T}\gamma_{a}[Y^{a},\theta'])={\rm Tr}\left(\psi^{T}\gamma^{i}(\psi J^{(1)}_{i}-J^{(2)}_{i}\psi)-r'\psi^{T}\gamma^{4}\psi\right),
\label{Fmass}
\end{equation}
which gives rise to the mass matrix of $\psi$. This, in turn, determines the zero-point energy of fermionic off-diagonal modes.
\subsubsection{A Field-theoretic Representation}
For $N$ becoming large, the quadratic terms in (\ref{Bmass1})-(\ref{Bmass5}) and in (\ref{Fmass}) can be drastically simplified under a field theory limit. Henceforth we proceed to define this field theory, following the arguments of \cite{AB}. 

First we represent the spin $j$ angular momentum $J_{i}$ in terms of certain differential operators. As usual, let us introduce the corresponding step operators $J_{\pm}=J_{1}\pm iJ_{2}$. Recall that there are two dielectric branes in our present system, whose configuration is described by (\ref{config}). On each such spherical 2-brane, we take angular coordinates $\vec{\vartheta} = (\vartheta,\varphi)$ with the volume element
\begin{equation}
d\Omega \;=\; \sin \vartheta\, d\vartheta\, d\varphi\;\equiv \;d^{2}\vec{\vartheta}.
\label{vol}
\end{equation}
The angular momentum associated to each brane can then be represented by
\begin{eqnarray}
J_+ &=& e^{i\varphi} ( \partial_\vartheta 
               + i \cot \vartheta \,\partial_\varphi)\nonumber\\
J_- &=&  -e^{-i\varphi} (\partial_\vartheta 
               - i\cot \vartheta\, \partial_\varphi)\nonumber\\
J_3 &=& -i \partial_\varphi.
\label{DiffOperator}
\end{eqnarray}

In this representation, the $N\times N$ matrix $A_{a}$ becomes a function of the angular coordinates $\vec{\vartheta}^{(1)}$ and $\vec{\vartheta}^{(2)}$, where the superscripts are used to distinguish the first and the second branes. Thus we get a field theory in 4+1 dimensions, and as a result the energy (or the Hamiltonian) of this theory depends on the differential operators (\ref{DiffOperator}) in a natural way.

Within the above field theory, one may ask how the finite $N$ effect (fuzziness) could be taken into account. For finite $N$ the matrix $A_{a}$ has two indices, each valued in a copy of the irreducible representation $j$. Thus, if we choose a basis $|j,m\rangle$ for each copy of the representation space, then $A_{a}$ can be written as a finite dimensional operator
\begin{equation}
A_{a}=\sum_{-j\leq m,\;m'\leq j}C_{mm'}|j,m\rangle^{(1)}\,\otimes\,\langle j,m'|^{(2)},
\label{finite}
\end{equation}
with $C_{mm'}$ being arbitrary coefficients\footnote{Similarly, the angular momentum operators $J^{(1,2)}_{i}$ acting on the first and the second copies of the representation space take the forms $J^{(1)}_{i}=\sum(J^{(1)}_{i})_{mm'}
|j,m\rangle^{(1)}\otimes\langle j,m'|^{(1)}$ and $J^{(2)}_{i}=\sum(J^{(2)}_{i})_{mm'}
|j,m\rangle^{(2)}\otimes\langle j,m'|^{(2)}$.}. In the field theory, one may represent the standard basis $|j,m\rangle^{(1,2)}$ by the spherical harmonic functions $\Psi_{jm}(\vec{\vartheta}^{(1,2)})$ of given $j=(N-1)/2$. For this representation (\ref{finite}) becomes
\begin{equation}
A_{a}(\vec{\vartheta}^{(1)},\vec{\vartheta}^{(2)})=\sum_{-j\leq m,\;m'\leq j}C_{mm'}\Psi_{jm}(\vec{\vartheta}^{(1)})\Psi_{jm'}(\vec{\vartheta}^{(2)})^{*}.
\label{FuncClass}
\end{equation}
Evidently, given fixed $N$, the space of such functions is finite dimensional, though the ``matrix indices'' $\vec{\vartheta}^{(1)}$, $\vec{\vartheta}^{(2)}$ are now continuous variables. This specifies what the allowed class of functions we should consider. The finite dimensional truncation (\ref{FuncClass}) imposed on the dynamical variables in our field theory is also compatible with the stringy exclusion principle.

As in \cite{AB}, let us establish rules for transforming the quadratic terms in (\ref{Bmass1})-(\ref{Bmass5}) and ({\ref{Fmass}) into the corresponding field theory expressions. Consider first multiplication of matrix blocks. The diagonal blocks are transformed into differential operators (\ref{DiffOperator}), so the product of two such blocks simply becomes the ordinary product of two differential operators. We thus have, for example, the correspondence
\begin{eqnarray}
J^{(1)}_{+}J^{(1)}_{+} &\rightarrow & \left[e^{i\varphi^{(1)}} \left( \partial_{\vartheta^{(1)}} 
               + i \cot \vartheta^{(1)} \partial_{\varphi^{(1)}}\right)\right]\cdot\left[e^{i\varphi^{(1)}} \left( \partial_{\vartheta^{(1)}} 
               + i \cot \vartheta^{(1)} \partial_{\varphi^{(1)}}\right)\right]\nonumber\\
J^{(2)}_{3}J^{(2)}_{-}&\rightarrow & \left[-i\partial_{\varphi^{(2)}}\right]\cdot\left[-e^{-i\varphi^{(2)}} \left( \partial_{\vartheta^{(2)}} 
               - i \cot \vartheta^{(2)} \partial_{\varphi^{(2)}}\right)\right].
\label{TransLaw1}
\end{eqnarray}
On the other hand, the off-diagonal blocks are transformed into functions of the form (\ref{FuncClass}), where $\vec{\vartheta}^{(1)}$ and $\vec{\vartheta}^{(2)}$ play the role of matrix indices. Accordingly, the product of an upper-right block and a lower-left block will become an integral, 
\begin{eqnarray}
A_{a}A_{b}^{\dag} & \rightarrow &\int d^{2}\vec{\vartheta}^{(2)} A_{a}(\vec{\vartheta}^{(1)}_{1},\vec{\vartheta}^{(2)})A^{*}_{b}(\vec{\vartheta}^{(2)},\vec{\vartheta}^{(1)}_{2}),\nonumber\\
A_{a}^{\dag}A_{b} & \rightarrow &\int d^{2}\vec{\vartheta}^{(1)} A^{*}_{a}(\vec{\vartheta}^{(2)}_{1},\vec{\vartheta}^{(1)})A_{b}(\vec{\vartheta}^{(1)},\vec{\vartheta}^{(2)}_{2}).
\label{TransLaw2}
\end{eqnarray}
Moreover, multiplying an off-diagonal block by a diagonal block corresponds naturally to the action of a differential operator on a function, {\it e.g.}, 
\begin{eqnarray}
J^{(1)}_{+}A_{a} &\rightarrow & e^{i\varphi^{(1)}} \left( \partial_{\vartheta^{(1)}} 
               + i \cot \vartheta^{(1)} \partial_{\varphi^{(1)}}\right)A_{a}(\vec{\vartheta}^{(1)},\vec{\vartheta}^{(2)}),\nonumber\\
J^{(2)}_{3}A^{\dag}_{a} &\rightarrow & -i\partial_{\varphi^{(2)}}A^{*}_{a}(\vec{\vartheta}^{(2)},\vec{\vartheta}^{(1)}).
\label{TransLaw3}
\end{eqnarray}

Next we are ready to give out rules for transforming a trace into a field theory expression. They are 
\begin{eqnarray}
{\rm Tr}(A_{a}\, O^{(2)} A^{\dag}_{b}\, O^{(1)}) & \rightarrow  & 
                              -\int A^{*}_{b}\, O^{(1)} O^{(2)} A_{a}\nonumber\\
{\rm Tr}(A_{a}\,A^{\dag}_{b}\, O^{(1)} O'^{(1)}) & \rightarrow &
                              \int A^{*}_{b}\, O^{(1)} O'^{(1)} A_{a}\nonumber\\
{\rm Tr}(A^{\dag}_{a}\,A_{b}\, O^{(2)} O'^{(2)}) & \rightarrow &
                              \int A^{*}_{a}\, O'^{(2)} O^{(2)} A_{b}
\label{TraceRule}
\end{eqnarray} 
where $O^{(1)}$ and $O^{(2)}$ denote the diagonal blocks belonging to the first and the second copies of the irreducible representation of spin $j$.

We hereby illustrate how the first relation is derived for $O^{(1)} = J^{(1)}_{+}$, 
$O^{(2)} =  J^{(2)}_{+}$ and we do it exactly the same way as in \cite{AB}. Let us substitute the trace by a sum over an orthonormal basis of the representation space. For concreteness, we shall choose spherical harmonic functions $\Psi_{jm}$ as the normalized basis. Now, each summation over an index corresponding to the first dielectric brane (located at $X^{4}\sim 0$) is replaced by an integral over $\vec{\vartheta}^{(1)}$, while each summation over an index corresponding to the second dielectric brane (located at $X^{4}\sim r$) is replaced by an integral over $\vec{\vartheta}^{(2)}$. The volume elements of these integrals are all defined as in (\ref{vol}). With the laws (\ref{TransLaw1})-(\ref{TransLaw3}) of multiplying matrix blocks, we have
\begin{eqnarray}
{\rm Tr}(A_{a}J^{(2)}_{+}A^{\dag}_{b}J^{(1)}_{+}) &\rightarrow &\sum_{m}\int d^{2}\vec{\vartheta}^{(1)}d^{2}\vec{\vartheta}^{(2)}d^{2}\vec{\vartheta'}^{(1)}\,\Psi_{jm}(\vec{\vartheta}^{(1)})^{*}A_{a}(\vec{\vartheta}^{(1)},\vec{\vartheta}^{(2)})\nonumber\\
&\times &\left[e^{i\varphi^{(2)}} \left( \partial_{\vartheta^{(2)}} 
               + i \cot \vartheta^{(2)} \partial_{\varphi^{(2)}}\right)\right]A^{*}_{b}(\vec{\vartheta}^{(2)},\vec{\vartheta'}^{(1)})\nonumber\\
&\times &\left[e^{i\varphi'^{(1)}} \left( \partial_{\vartheta'^{(1)}} 
               + i \cot \vartheta'^{(1)} \partial_{\varphi'^{(1)}}\right)\right]\Psi_{jm}(\vec{\vartheta'}^{(1)}).
\label{tmp}
\end{eqnarray}
This expression can be simplified with the following two observations. 

First, given two arbitrary functions $F(\vec{\vartheta})$ and $G(\vec{\vartheta})$ one has, by integration by parts,
\begin{equation}
\int d^{2}\vec{\vartheta}\,F(\vec{\vartheta})\,e^{i\varphi}(\partial_{\vartheta}+i\cot \vartheta\,\partial_{\varphi})\,G(\vec{\vartheta})=-\int d^{2}\vec{\vartheta}\left[e^{i\varphi}(\partial_{\vartheta}+i\cot \vartheta\,\partial_{\varphi})F(\vec{\vartheta})\right]G(\vec{\vartheta}).
\label{tmp1}
\end{equation}
This could be understood as $\langle \overline{F},J_{+}G\rangle=-\langle \overline{J_{+}F},G\rangle$, which is a consequence of $(J_{+})^{\dag}=J_{-}=-\bar{J}_{+}$ (a property valid in the representation (\ref{DiffOperator})).

Second, the sum in (\ref{tmp}) was originally performed over $m=-j,-j+1,\cdots,j$ with fixed $j=(N-1)/2$, but one can extend it to a sum running over a set of
$j'$'s instead of sticking to the given one, $j' = j$. The point is that in (\ref{tmp}), the function $A_{a}(\vec{\vartheta}^{(1)},\vec{\vartheta}^{(2)})$ takes the form (\ref{FuncClass}), which is orthogonal to the spherical harmonic functions $\Psi_{j'm'}(\vec{\vartheta}^{(1)})$ for any $j'\neq j$. In other words, we have
\begin{eqnarray}
& & \sum_{j',\,m'}\int d^{2}\vec{\vartheta}^{(1)}\,\Psi_{j'm'}(\vec{\vartheta}^{(1)})^{*}A_{a}(\vec{\vartheta}^{(1)},\vec{\vartheta}^{(2)})F_{j'm'}(\vec{\vartheta}^{(2)},\vec{\vartheta'}^{(1)})\nonumber\\
&=&\sum_{m}\int d^{2}\vec{\vartheta}^{(1)}\,\Psi_{jm}(\vec{\vartheta}^{(1)})^{*}A_{a}(\vec{\vartheta}^{(1)},\vec{\vartheta}^{(2)})F_{jm}(\vec{\vartheta}^{(2)},\vec{\vartheta'}^{(1)}),
\label{tmp2}
\end{eqnarray}
so when the sum in (\ref{tmp}) is extended to $\sum_{j'm'}$, only those terms with $j'=j$ will contribute.

Thus, after replacing $\sum_{m}$ by $\sum_{j'm'}$, we can rewrite the right hand side of (\ref{tmp}) as
\begin{eqnarray}
& &\int d^{2}\vec{\vartheta}^{(1)}d^{2}\vec{\vartheta}^{(2)}
\left[e^{i\varphi^{(2)}} \left( \partial_{\vartheta^{(2)}} 
               + i \cot \vartheta^{(2)} \partial_{\varphi^{(2)}}\right)A_{a}(\vec{\vartheta}^{(1)},\vec{\vartheta}^{(2)})\right]\nonumber\\
&\times &\left[e^{i\varphi^{(1)}} \left( \partial_{\vartheta^{(1)}} 
               + i \cot \vartheta^{(1)} \partial_{\varphi^{(1)}}\right)A^{*}_{b}(\vec{\vartheta}^{(2)},\vec{\vartheta}^{(1)})\right]\nonumber\\
&=&-\int\left[e^{i\varphi^{(1)}} \left( \partial_{\vartheta^{(1)}} 
               + i \cot \vartheta^{(1)} \partial_{\varphi^{(1)}}\right)e^{i\varphi^{(2)}} \left( \partial_{\vartheta^{(2)}} 
               + i \cot \vartheta^{(2)} \partial_{\varphi^{(2)}}\right)A_{a}(\vec{\vartheta}^{(1)},\vec{\vartheta}^{(2)})\right]\nonumber\\
&\times&A^{*}_{b}(\vec{\vartheta}^{(2)},\vec{\vartheta}^{(1)})\;=\;-\int A^{*}_{b}J^{(1)}_{+}J^{(2)}_{+}A_{a}.
\label{tmp3}
\end{eqnarray}
Here we have used the relation
$\sum_{j'm'}\Psi_{j'm'}(\vec{\vartheta})^{*}\Psi_{j'm'}(\vec{\vartheta'})=\delta(\vec{\vartheta}-\vec{\vartheta'})$ and applied
(\ref{tmp1}) several times. This completes our derivation of ${\rm Tr}(A_{a}J^{(2)}_{+}A^{\dag}_{b}J^{(1)}_{+}) \rightarrow -\int A^{*}_{b}J^{(1)}_{+}J^{(2)}_{+}A_{a}$.

Having established these transformation rules, we could at last write down the field theory expressions of (\ref{Bmass1})-(\ref{Bmass5}) and (\ref{Fmass}). The results simply read
\begin{eqnarray}
\sum_{1\leq i,j\leq 3}{\rm Tr}[Y^{i},Y^{j}]^{2}&=&4\int\left(-A^{*}_{i}\hat{J}^{2}A_{i}+A^{*}_{i}\hat{J}_{j}\hat{J}_{i}A_{j}+A^{*}_{i}[\hat{J}_{j},\hat{J}_{i}]A_{j}\right),\nonumber\\
\sum_{i=1,\,2,\,3}{\rm Tr}[Y^{i},Y^{4}]^{2} &=& -2\int\left(A^{*}_{4}\hat{J}^{2}A_{4}+r'A^{*}_{i}\hat{J}_{i}A_{4}+r'A^{*}_{4}\hat{J}_{i}A_{i}+r'^{2}A^{*}_{i}A_{i}\right),\nonumber\\
\sum_{i\leq 3,\;\mu\geq 5}{\rm Tr}[Y^{i},Y^{\mu}]^{2} &=& -2\int A^{*}_{\mu}\hat{J}^{2}A_{\mu},\nonumber\\
\sum_{\mu=5,\cdots,9}{\rm Tr}[Y^{4},Y^{\mu}]^{2} &=& -2\int r'^{2}A^{*}_{\mu}A_{\mu},\nonumber\\
{\rm Tr}\,(Y^{i}\,Y^{j}\,Y^{k})\,\epsilon_{ijk} &=& 3\int A^{*}_{i}\hat{J}_{j}A_{k}\,\epsilon_{ijk},\nonumber\\
{\rm Tr}\,(\theta'^{T}\gamma_{a}[Y^{a},\,\theta']) &=&-\int\psi^{T}(\gamma^{i}\hat{J}_{i}+r'\gamma^{4})\psi
\label{field}
\end{eqnarray}
where
\begin{equation}
\hat{J}_{i}\equiv J^{(1)}_{i}+J^{(2)}_{i}
\label{tot}
\end{equation}
is the total angular momentum. 

Clearly, the total spin of our field-theoretic system takes different values $\hat{j}=0,1,\cdots,2j\equiv N-1$, corresponding to various irreducible components of $j\otimes j$. The representation space of $j\otimes j$ is isomorphic to the function space spanned by (\ref{FuncClass}), so in this system the differential operator $\hat{J}^{2}$ has only finitely many eigenvalues $\hat{j}(\hat{j}+1)$ (with degeneracy $2\hat{j}+1$), each associated with an irreducible component of $j\otimes j$.
\subsubsection{Mass Matrices and Their Spectra}
Let us divide the matrix model energy (\ref{energy1}) into a bosonic part $V_{B}$ and a fermionic part $V_{F}$, and perform a transformation to get the corresponding field theory expressions. With the help of (\ref{field}), one sees that the bosonic energy takes a form
\begin{eqnarray}
V_{B} &=&-\frac{T_{0}}{4\lambda^{2}}\sum_{a\neq b}{\rm Tr}[X^{a},X^{b}]^{2}
+\frac{2iT_{0}f}{3\lambda^{2}}{\rm Tr}(X^{i}X^{j}X^{k})\epsilon_{ijk}\nonumber\\
&=&-\frac{T_{0}f^{4}}{4\lambda^{2}}\sum_{a\neq b}{\rm Tr}[Y^{a},Y^{b}]^{2}
+\frac{2iT_{0}f^{4}}{3\lambda^{2}}{\rm Tr}(Y^{i}Y^{j}Y^{k})\epsilon_{ijk}\nonumber\\
&=&\frac{T_{0}f^{4}}{\lambda^{2}}\int\left[A^{*}_{\mu}(\hat{J}^{2}+r'^{2})A_{\mu}+{\bf A}^{\dag}{\cal M}_{B}^{2}{\bf A}\right],
\label{energyB}
\end{eqnarray}
where ${\bf A}$ is an assembly of the first four bosonic off-diagonal modes, and ${\cal M}_{B}$ denotes their mass matrix:
\begin{equation}
{\bf A}=\pmatrix{A_{1} \cr A_{2} \cr A_{3} \cr A_{4}},\quad
{\cal M}_{B}^{2}=\hat{J}^{2}+r'^{2}-\pmatrix{\hat{J}_{1}^{2} & \hat{J}_{1}\hat{J}_{2} & \hat{J}_{1}\hat{J}_{3} & -r'\hat{J}_{1} \cr
\hat{J}_{2}\hat{J}_{1} & \hat{J}_{2}^{2} & \hat{J}_{2}\hat{J}_{3} & -r'\hat{J}_{2}\cr
\hat{J}_{3}\hat{J}_{1} & \hat{J}_{3}\hat{J}_{2} & \hat{J}_{3}^{2} & -r'\hat{J}_{3} \cr
-r'\hat{J}_{1} & -r'\hat{J}_{2} & -r'\hat{J}_{3} & r'^{2}}.
\label{MMB}
\end{equation}
Obviously, the four by four operator-valued matrix in (\ref{MMB}) is hermitian.

In a similar way, we can write down the fermionic energy
\begin{eqnarray}
V_{F} &=& -\frac{T_{0}}{\lambda}\,{\rm Tr}\,(\theta^{T}\gamma_{a}\,[X^{a},\,\theta])\nonumber\\
&=& -\frac{f}{\lambda}\,{\rm Tr}\,(\theta'^{T}\gamma_{a}\,[Y^{a},\,\theta'])\nonumber\\
&=&\frac{f}{\lambda}\int \psi^{T}{\cal M}_{F}\,\psi
\label{energyF}
\end{eqnarray}
with a mass matrix ${\cal M}_{F}$, whose explicit expression is given by:
\begin{equation}
{\cal M}_{F}=\gamma_{1}\hat{J}_{1}+\gamma_{2}\hat{J}_{2}+\gamma_{3}\hat{J}_{3}+r'\gamma_{4}.
\label{MMF}
\end{equation}

As mentioned before, the zero-point energy for the off-diagonal modes $(A_{a}$, $\psi)$ comes from the mass spectra of ${\cal M}_{B}$ and ${\cal M}_{F}$, as well as from the masses of $A_{\mu}$, $\mu\geq 5$. Let us give now a detailed discussion of these mass spectra.

We set out to calculate the fermionic spectrum first. From (\ref{MMF}), one has
\begin{eqnarray}
{\cal M}_{F}^{2} &=& \hat{J}^{2}_{1} + \hat{J}^{2}_{2} + \hat{J}^{2}_{3}+ r'^{2} + i\gamma_{1} \gamma_{2} \hat{J}_{3} 
    + i\gamma_{2} \gamma_{3} \hat{J}_{1} + i\gamma_{3} \gamma_{1} \hat{J}_{2}\nonumber\\
      &\equiv& \hat{J}^{2} + r'^{2} + {\cal N}_{F},
\label{FM}
\end{eqnarray}
so that the eigenvalues of ${\cal M}_{F}^{2}$ are decomposed into  $\hat{j}(\hat{j}+1)+r'^{2}$ (with $0\leq\hat{j}\leq N-1$) plus the eigenvalues of ${\cal N}_{F}$. Furthermore, since
\begin{eqnarray}
{\cal N}_{F}^{2} &=& (i\gamma_{2} \gamma_{3} \hat{J}_{1} +
                                 i\gamma_{3} \gamma_{1} \hat{J}_{2} +
                                 i\gamma_{1} \gamma_{2} \hat{J}_{3})^2\nonumber\\
                            &=& \hat{J}^{2}_{1} + \hat{J}^{2}_{2} + {J}^{2}_{3} +    i\gamma_{1} \gamma_{2} \hat{J}_{3} + 
                                 i\gamma_{3} \gamma_{1} \hat{J}_{2} + 
                                 i\gamma_{2} \gamma_{3} \hat{J}_{1}\nonumber\\
                            &=& \hat{J}^{2} + {\cal N}_{F},\nonumber                                
\end{eqnarray}
we find that half of the eigenvalues of ${\cal N}_{F}$ take the value $\hat{j}+1$, and half take $-\hat{j}$. As a result, the fermionic spectrum is given by
\begin{equation}
\omega_{F}^{2}=\frac{f^{2}}{\lambda^{2}}\times
\left\{
             \begin{array}{l}
                 (\hat{j}+1)^{2} + r'^{2}   \\             
                 \hat{j}^{2} + r'^{2}
             \end{array}
          \right.
\label{specF}
\end{equation}
of which each is $4\times (2\hat{j}+1)$-fold degenerate.

Next we consider the spectrum of the bosonic modes $A_{\mu}$, $\mu=5,6,7,8,9$. The mass matrix could literately be read off from the first term in (\ref{energyB}), which takes a diagonal form and whose elements have the values\footnote{The coefficient $f/\lambda$ in the frequency $\omega_{B}$ comes from the Hamiltonian ${\cal H}\sim\Pi^{2}_{\mu}/(T_{0}f^{2})+(T_{0}f^{4}/\lambda^{2})A^{\dag}_{\mu}A_{\mu}$ for the oscillator $A_{\mu}$, where $\Pi_{\mu}=T_{0}f^{2}\dot{A}_{\mu}$ is the canonical momentum conjugate to $A_{\mu}$.}
\begin{equation}
\omega_{B}^{2}=\frac{f^{2}}{\lambda^{2}}\left[\hat{j}(\hat{j}+1)+r'^{2}\right].
\label{specB1}
\end{equation}
Thus, for each fixed $\hat{j}$, we have totally $5\times (2\hat{j}+1)$ independent states corresponding to the eigenvalue (\ref{specB1}), all of which are degenerate.

We finally come to the mass spectrum of the remaining four bosonic modes, $A_{1},A_{2},A_{3},A_{4}$. This requires us to diagonalize the operator-valued matrix ${\cal M}_{B}$. For this purpose, let us take the unitary matrix
\begin{equation}
U= \pmatrix{ 
  \frac{1}{\sqrt{2}} & -\frac{1}{\sqrt{2}} & 0 &0 \cr
  \frac{i}{\sqrt{2}} & \frac{i}{\sqrt{2}} &0 & 0 \cr 
  0 & 0 & 1 & 0\cr
  0 & 0 & 0 & 1}
\label{unitary}
\end{equation}
as in \cite{AB} and consider the conjugate of ${\cal M}_{B}$:
\begin{equation}
\widetilde{\cal M}_{B}^{2}\equiv U^{\dag}{\cal M}_{B}^{2}U=\hat{J}^{2}+r'^{2}-{\cal N}_{B}
\label{conjugate}
\end{equation}
where ${\cal N}_{B}$ is computed from (\ref{MMB}) and (\ref{unitary}). Explicitly we have
\begin{equation}
{\cal N}_{B}=\pmatrix{\frac{1}{2}\hat{J}_{-}\hat{J}_{+} & -\frac{1}{2}\hat{J}_{-}^{2} & \frac{1}{\sqrt{2}}\hat{J}_{-}\hat{J}_{3} & -\frac{1}{\sqrt{2}}r'\hat{J}_{-}\cr
-\frac{1}{2}\hat{J}_{+}^{2} & \frac{1}{2}\hat{J}_{+}\hat{J}_{-} & -\frac{1}{\sqrt{2}}\hat{J}_{+}\hat{J}_{3} & \frac{1}{\sqrt{2}}r'\hat{J}_{+}\cr
\frac{1}{\sqrt{2}}\hat{J}_{3}\hat{J}_{+} & -\frac{1}{\sqrt{2}}\hat{J}_{3}\hat{J}_{-} & \hat{J}_{3}^{2} & -r'\hat{J}_{3}\cr
-\frac{1}{\sqrt{2}}r'\hat{J}_{+} & \frac{1}{\sqrt{2}}r'\hat{J}_{-} & -r'\hat{J}_{3} & r'^{2}}.
\label{NMB}
\end{equation}

Since the matrix (\ref{NMB}) is still far from being diagonalized, one needs further to solve the eigenequation
\begin{equation}
{\cal N}_{B}\, v=\Lambda v.
\label{eigenEqn}
\end{equation}
The solutions to (\ref{eigenEqn}) can be constructed along the following line. By inspection, one finds that the eigenvector $v$ takes the form
\begin{equation}
v=v_{\hat{j},\,\hat{m}}\equiv\pmatrix{\alpha_{1}\;\Psi_{\hat{j},\;\hat{m}}\,(\vec{\vartheta})\cr\alpha_{2}\Psi_{\hat{j},\,\hat{m}+2}(\vec{\vartheta})\cr\alpha_{3}\Psi_{\hat{j},\,\hat{m}+1}(\vec{\vartheta})\cr\alpha_{4}\Psi_{\hat{j},\,\hat{m}+1}(\vec{\vartheta})}
\label{eigenVect}
\end{equation}
with numerical coefficients $\alpha_{1},\alpha_{2},\alpha_{3},\alpha_{4}$. Hence (\ref{eigenEqn}) reduces to an algebraic eigenequation
\begin{equation}
\pmatrix{a_{1}b_{1}-\Lambda & a_{1}b_{2} & a_{1}b_{3} & a_{1}b_{4}\cr
a_{2}b_{1} & a_{2}b_{2}-\Lambda & a_{2}b_{3} & a_{2}b_{4}\cr
a_{3}b_{1} & a_{3}b_{2} & a_{3}b_{3}-\Lambda & a_{3}b_{4}\cr
b_{1} & b_{2} & b_{3} & b_{4}-\Lambda}\pmatrix{\alpha_{1}\cr
\alpha_{2}\cr \alpha_{3}\cr\alpha_{4}}=0,
\label{alg}
\end{equation}
where
\begin{eqnarray}
a_{1} &=& -\frac{1}{\sqrt{2}r'}[(\hat{j}-\hat{m})(\hat{j}+\hat{m}+1)]^{1/2}\nonumber\\
a_{2} &=& \frac{1}{\sqrt{2}r'}[(\hat{j}-\hat{m}-1)(\hat{j}+\hat{m}+2)]^{1/2}\nonumber\\
a_{3} &=& -\frac{1}{r'}(\hat{m}+1)\nonumber\\
b_{1} &=& -\frac{r'}{\sqrt{2}} [(\hat{j}-\hat{m})(\hat{j}+\hat{m} +1)]^{1/2}\nonumber\\ 
b_{2} &=&\frac{r'}{\sqrt{2}} [(\hat{j}+\hat{m}+2)(\hat{j}-\hat{m}-1)]^{1/2} \nonumber\\
b_{3} &=& -r'(\hat{m}+1)\nonumber\\
b_{4} &=& r'^2.
\label{algCoeff}
\end{eqnarray}
One may check that the $4\times 4$ matrix in (\ref{alg}) is symmetric, indicating that our new eigenvalue problem is valid. This problem looks much simpler than the original one, (\ref{eigenEqn}), and can be solved explicitly. Each solution $(\Lambda,\alpha_{1},\alpha_{2},\alpha_{3},\alpha_{4})$ to (\ref{alg}) then determines a set of (degenerate) solutions $(\Lambda, v_{\hat{j},\,\hat{m}})$ of the original eigenequation.

Now given a pair of fixed quantum numbers $\hat{j}, \hat{m}$, it is fairly easy to work out the eigenvalues $\Lambda$ in (\ref{alg}) as
\begin{equation}
\Lambda = \left\{
                \begin{array}{ll}
                   0, & \hbox{3-fold degenerate;}\\
                   \hat{j}(\hat{j}+1)+r'^{2},  \quad\quad& \hbox{singlet.}
                \end{array}                  
\right.
\label{eigenValue}
\end{equation}
Note that these values may depend on $\hat{j}$, but they are independent of $\hat{m}$. Thus, modulo some subtleties which we shall discuss in a moment, one concludes that there are totally $3\times (2\hat{j}+1)$ degenerate states corresponding to the eigenvalue $\Lambda=0$, and totally $(2\hat{j}+1)$ degenerate states corresponding to $\Lambda=\hat{j}(\hat{j}+1)+r'^{2}$. From (\ref{conjugate}) we see that the eigenvalues of $\widetilde{\cal M}_{B}^{2}$ are given by $\hat{j}(\hat{j}+1)+r'^{2}-\Lambda$. This allows us to write down the mass spectrum for the bosonic off-diagonal modes $A_{1},A_{2},A_{3},A_{4}$,
\begin{equation}
\tilde{\omega}_{B}^{2}=\frac{f^{2}}{\lambda^{2}}\times\left\{
                \begin{array}{ll}
                   \hat{j}(\hat{j}+1)+r'^{2}\\
                    0
                \end{array}                  
\right.
\label{specB2}
\end{equation}
of which the first eigenvalue is $3\times(2\hat{j}+1)$-fold degenerate\footnote{The degeneracy of the second eigenvalue is $2\hat{j}+1$. Since the frequency vanishes, the associated modes do not contribute to the zero-point energy; they correspond to certain gauge transformations \cite{AB}.}.

The above discussion rests implicitly on an assumption, namely for $\hat{j}$  held fixed while $\hat{m}$ running from $-\hat{j}$ to $\hat{j}$, the eigenvectors $v_{\hat{j},\,\hat{m}}$ associated to a given $\Lambda$ form a set of $(2\hat{j}+1)n_{\hat{j}}$ linearly independent states, here $n_{\hat{j}}$ is the degeneracy of $(\alpha_{1},\alpha_{2},\alpha_{3},\alpha_{4}, \Lambda)$ which solves (\ref{alg}). (According (\ref{eigenValue}), we have $n_{\hat{j}}=3$ for $\Lambda=0$ and $n_{\hat{j}}=1$ for $\Lambda=\hat{j}(\hat{j}+1)+r'^{2}$.) Actually, for $\Lambda=0$ the eigenequation (\ref{alg}) gives the following relation
\begin{equation}
b_{1} \alpha_{1} + b_{2} \alpha_{2} + b_{3} \alpha_{3} +b_{4} \alpha_{4} = 0.
\label{eigen}
\end{equation}
So fixing a generic $\hat{m}$ we have three independent states 
\begin{eqnarray}
v^{(1)}_{\hat{j},\,\hat{m}}&=&\pmatrix{[(\hat{j}+\hat{m} 		  +2)(\hat{j}-\hat{m}-1)]^{1/2} \Psi_{\hat{j},\,\hat{m}}\cr 
               [(\hat{j}-\hat{m})(\hat{j}+\hat{m}+1)]^{1/2} \Psi_{\hat{j},\,\hat{m}+2}\cr
               0\cr
               0}\nonumber  \\
v^{(2)}_{\hat{j},\,\hat{m}}&=&\pmatrix{(\hat{m} +1)\Psi_{\hat{j},\,\hat{m}}\cr 
               0\cr
            -\frac{1}{\sqrt{2}}[(\hat{j}-\hat{m})(\hat{j}+\hat{m}+1)]^{1/2}\Psi_{\hat{j},\,\hat{m}+1}\cr
               0}\nonumber\\
v^{(3)}_{\hat{j},\,\hat{m}}&=&\pmatrix{r'\Psi_{\hat{j},\,\hat{m}}\cr 
               0\cr
               0\cr
            -\frac{1}{\sqrt{2}}[(\hat{j}-\hat{m})(\hat{j}+\hat{m}+1)]^{1/2}\Psi_{\hat{j},\,\hat{m}+1}}
\label{v123}
\end{eqnarray}       
and, for $\hat{m}$ running from $-\hat{j}$ to $\hat{j}$, the total number of such states is expected to be $3\times (2\hat{j}+1)$. But there are some subtleties here: if  $\hat{m}$ takes the special value $\hat{j}-1$, then  $v^{(1)}_{\hat{j},\,\hat{m}}$ vanishes and we get only two eigenvectors $v^{(2)}_{\hat{j},\,\hat{m}}$, $v^{(3)}_{\hat{j},\,\hat{m}}$. Moreover, for $\hat{m}=\hat{j}$ the three eigenstates in (\ref{v123}) even coincide! Put together it would appear that there are three states missing. 

To resolve this problem, we extend the allowed values of $\hat{m}$ from $-\hat{j}\leq\hat{m}\leq\hat{j}$ to $-\hat{j}-2\leq\hat{m}\leq\hat{j}$ by requiring $\Psi_{\hat{j},\,-\hat{j}-1}=\Psi_{\hat{j},\,-\hat{j}-2}=0$, and  construct some new states $v_{\hat{j},\,-\hat{j}-1}$, $v_{\hat{j},\,-\hat{j}-2}$ as in (\ref{eigenVect}). Such states become extra eigenvectors provided their coefficients $\alpha_{1},\cdots\alpha_{4}$ satisfy (\ref{alg}) or, in the case of $\Lambda=0$, (\ref{eigen}). For $\hat{m}= -\hat{j}-1$ we have $b_{1} =0$, so there are two independent eigenstates determined by 
$$
b_{2} \alpha_{2} + b_{3} \alpha_{3} + b_{4} \alpha_{4}= 0.
$$
In addition, we have one more eigenstate arising from $\hat{m}= -\hat{j}-2$, which is simply 
$$
v_{\hat{j},\,-\hat{j}-2}=\pmatrix{0\cr\psi_{\hat{j},\,-\hat{j}}\cr 0\cr 0}.
$$
Everything included we still find $3\times(2\hat{j}+1)$ degenerate eigenstates coming from the sector of $(\Lambda,\hat{J}^{2}) = (0,\hat{j}(\hat{j}+1))$. A similar argument applies to the sector of $(\Lambda,\hat{J}^{2}) =(\hat{j}(\hat{j}+1)+r'^{2},\hat{j}(\hat{j}+1))$, where we have totally $2\hat{j}+1$ independent eigenstates.
\subsubsection{The Zero-point Energy}
The zero-point energy is now computed from the sum of all the bosonic frequencies $(\omega_{B},\tilde{\omega}_{B})$, minus the sum of all the fermionic frequencies $\omega_{F}$, with each frequency weighted by its degeneracy. We have seen that the degeneracies for $\omega_{B}$, $\tilde{\omega}_{B}$, and the two $\omega_{F}$'s in (\ref{specF}) are $5(2\hat{j}+1)$, $3(2\hat{j}+1)$, $4(2\hat{j}+1)$, $4(2\hat{j}+1)$, respectively, so the total zero-point energy reads
\begin{eqnarray}
V(r) &=& \frac{f}{\lambda}\,\sum^{N-1}_{\hat{j}=0}(2\hat{j}+1)\left\{
8\sqrt{\hat{j}(\hat{j}+1) + r'^2} - 4\sqrt{(\hat{j}+1)^2 + r'^2}- 4\sqrt{\hat{j}^2 + r'^2} \right\}\nonumber\\
&=& \frac{2f}{\lambda}\,\sum^{N-1}_{\hat{j}=0}(2\hat{j}+1) (-\frac{1}{r'})\;=\; -\frac{2N^{2}f^{2}}{\lambda\, r}+{\cal O}(\frac{1}{r^{2}}).
\label{ZeroEnergy1}
\end{eqnarray}
This gives the $1/r$-law of interactions mentioned in the introduction. Note that the factor $f^{2}$ appeared in this two-body potential agrees exactly with our previous estimation (\ref{1/r}) using the Viral theorem.

Let us end with an identification of various quantities in (\ref{generic}) for the current system. We have nine bosons $A_{a}$, one of which has vanishing frequencies and thus does not contribute to the zero-point energy. So effectively we find $n_{b}=8$. The number(s) $\bf k$ labelling quantized modes is simply the total angular momentum $\hat{j}$ and, for each boson, the coefficients $a_{i}({\bf k})$, $b_{i}({\bf k})$ in (\ref{generic}) are given by the degeneracy $2\hat{j}+1$ and the Casimir $\hat{j}(\hat{j}+1)$, respectively:
\begin{equation}
a_{i}({\bf k})=\frac{1}{\lambda}\,(2\hat{j}+1),\quad
b_{i}({\bf k})=f^{2}\,\hat{j}(\hat{j}+1);\quad 1\leq i\leq n_{b}=8.
\label{ab}
\end{equation}
Also, there are totally $2^{\left[\frac{9}{2}\right]}=16$ fermions $\theta$ in the system, half of which can be identified with creation operators, and half with annihilation operators. The effective number of fermions $\psi$ is thus the same as bosons: $n_{f}=8$. According to (\ref{specF}), these fermions can be divided into two groups, each contains four and has a different mass spectrum. Hence the coefficients $c_{i}({\bf k})$, $d_{i}({\bf k})$ in (\ref{generic}) are computed by
\begin{equation}
c_{i}({\bf k})=\frac{1}{\lambda}\,(2\hat{j}+1),\quad
d_{i}({\bf k})=f^{2}\times\left\{
                \begin{array}{ll}
                   (\hat{j}+1)^{2},\quad  & \hbox{for $i=1,3,5,7$;}\\
                    \hat{j}^{2}, & \hbox{for $i=2,4,6,8$.}
                \end{array}                  
\right.
\label{cd}
\end{equation}
Clearly, for each boson-fermion pair, we have $a_{i}({\bf k})=c_{i}({\bf k})$ but $b_{i}({\bf k})\neq d_{i}({\bf k})$. This explicitly shows that SUSY is lost in our system. The $1/r$ behavior of the potential then follows directly from the identity $a_{i}({\bf k})=c_{i}({\bf k})$.
\section{Incorporation of Mass Terms} \label{s3}
From the point of view of the AdS/CFT correspondence, we can add some mass terms to the action of dielectric 5-branes and construct an ${\cal N}=1^{*}$ supersymmetric theory, which describes certain gauge fields in the confining phase \cite{PS}. Motivated by this, we will now incorporate the mass terms
\begin{equation}
V_{M}=\frac{T_{0}M^{2}}{2}\sum_{i=1,2,3}{\rm Tr}\,(X^{i})^{2}
\label{mass}
\end{equation}
in our system of dielectric 2-branes, and study how this can influence the long-range two-body potential. In particular, we wish to test whether there exist certain additional cancellations between bosonic and fermionic fields due, possibly, to the partial restoration of supersymmetry. Our discussion is a slight modification of that presented in Section~\ref{s2}.

We begin with the bosonic part of the total energy:
\begin{equation}
V_{B}=-\frac{T_{0}}{4\lambda^{2}}\sum_{a\neq b}{\rm Tr}[X^{a},X^{b}]^{2}
+\frac{2ifT_{0}}{3\lambda^{2}}{\rm Tr}(X^{i}X^{j}X^{k})\epsilon_{ijk}+\frac{T_{0}M^{2}}{2}\sum_{i=1,2,3}{\rm Tr}(X^{i})^{2}.
\label{energy2}
\end{equation}
From this expression, it is straightforward to derive the static equations of motion,
\begin{equation}
[X^{j},[X^{i},X^{j}]]-if\epsilon_{ijk}[X^{j},X^{k}]-\lambda^{2}M^{2}X^{i}=0.
\label{EOM2}
\end{equation}
The above equations of motion are solved by $X^{i}=\beta J^{i}$ with $-2\beta^{2}+2\beta f -M^{2}\lambda^{2}=0$, so we have, classically,
\begin{equation}
X^{i}=\beta J^{i},\quad\beta\equiv\frac{f+\sqrt{f^{2}-2M^{2}\lambda^{2}}}{2}.
\end{equation}
Here we choose only one root of $\beta$ such that it tends to $f$ (rather than 0) when the mass $M$ vanishes.

As in (\ref{noM}), we then rescale $X^{a}$, $r$, $\theta$ as
\begin{equation}
X^{a}=\beta Y^{a}, \quad r=\beta r', \quad\theta=T_{0}^{-1/2}\theta'
\label{noM1}
\end{equation}
with $a=(i,4,\mu)_{1\leq i\leq 3,\;5\leq\mu\leq 9}$, and make the ansatz
\begin{eqnarray}
Y^{i}=\pmatrix{J^{(1)}_{i} & A_{i}\cr
A_{i}^{\dag} & J^{(2)}_{i}},\quad Y^{4}=\pmatrix{0 & A_{4}\cr
A_{4}^{\dag} & r'},\quad Y^{\mu}=\pmatrix{0 & A_{\mu}\cr
A_{\mu}^{\dag} & 0}.
\label{bFluc1}
\end{eqnarray}
Substituting these into the bosonic potential, we may expand $V_{B}$ in powers of the off-diagonal modes $A_{a}$. 

The leading terms (${\cal O}(A^{0})$) in the potential give rise to the classical energy of the configuration, which is computed by:
\begin{equation}
V^{(0)}_{B}=\frac{T_{0}C_{2}Nf^{2}\beta^{2}}{\lambda^{2}}\left(\frac{M^{2}\lambda^{2}}{f^{2}}-\frac{4}{9}\right)\left(\frac{1}{2}+\frac{1}{1+3\sqrt{1-\frac{2M^{2}\lambda^{2}}{f^{2}}}}\right)
\label{ClassicEngergy}
\end{equation}
where $C_{2}$ denotes, as before, the quadratic Casimir of the representation $J^{(1,2)}_{i}$. Thus, for the $N$-dimensional irreducible representation, we have $C_{2}=\frac{N^{2}-1}{4}$. 

Let us make some simple observations concerning the classical energy (\ref{ClassicEngergy}).

(1) If 
\begin{equation}
\frac{M^{2}\lambda^{2}}{f^{2}}>\frac{1}{2},
\end{equation}
then the critical value of the classical potential becomes complex, indicating that the system is classically unstable and the potential has no real minimal values. Thus, the dielectric effect can be seen only for the mass $M$ not being too large.

(2) When
\begin{equation}
\frac{4}{9}\leq\frac{M^{2}\lambda^{2}}{f^{2}}\leq\frac{1}{2},
\end{equation}
the potential is non-negative. The limiting value 
\begin{equation}
\frac{M^{2}\lambda^{2}}{f^{2}}=\frac{4}{9}
\label{SUSY}
\end{equation}
of the mass corresponds to vanishing of the classical energy, at which SUSY is supposed to be partially restored.

(3) Finally, when
\begin{equation}
0\leq \frac{M^{2}\lambda^{2}}{f^{2}} < \frac{4}{9},
\end{equation}
the potential is negative. The limiting value $\frac{M^{2}\lambda^{2}}{f^{2}}=0$ corresponds to the case we considered in Section~\ref{s2}, where $\beta=f$ and the classical energy (\ref{ClassicEngergy}) simply reduces to (\ref{2Npotential-reducible}).

Now, the terms next to the leading ones in the potential are those quadratic in $A_{a}$. By replacing the matrix-traces with the corresponding field theory expressions, we find that the quadratic part of the bosonic potential takes the form:
\begin{equation}
V_{B}^{(2)}=\frac{T_{0}\beta^{4}}{\lambda^{2}}\int\left[\sum_{\mu=5}^{9}A_{\mu}^{\dag}(\hat{J}^{2}+r'^{2})A_{\mu}+{\bf A}^{\dag}{\cal M}_{B}^{2}{\bf A}\right],
\label{energyB1}
\end{equation}
where $\hat{J}_{i}=J^{(1)}_{i}+J^{(2)}_{i}$ is the total angular momentum,
\begin{eqnarray}
{\bf A}=\pmatrix{A_{1}\cr A_{2}\cr A_{3}\cr A_{4}},\quad
{\cal M}_{B}^{2}=\hat{J}^{2}+r'^2+\frac{M^{2}\lambda^{2}}{\beta^{2}}-{\cal N}_{B},
\end{eqnarray}
and ${\cal N}_{B}$ the operator-valued matrix
\begin{eqnarray}
\pmatrix{(\hat{J}_{1})^{2} & (\frac{2f}{\beta}-1)[\hat{J}_{1},\hat{J}_{2}]+\hat{J}_{2}\hat{J}_{1} & (\frac{2f}{\beta}-1)[\hat{J}_{1},\hat{J}_{3}]+\hat{J}_{3}\hat{J}_{1} & -r'\hat{J}_{1}\cr
(\frac{2f}{\beta}-1)[\hat{J}_{2},\hat{J}_{1}]+\hat{J}_{1}\hat{J}_{2} & (\hat{J}_{2})^{2} & (\frac{2f}{\beta}-1)[\hat{J}_{2},\hat{J}_{3}]+\hat{J}_{3}\hat{J}_{2} & -r'\hat{J}_{2}\cr
(\frac{2f}{\beta}-1)[\hat{J}_{3},\hat{J}_{1}]+\hat{J}_{1}\hat{J}_{3} & (\frac{2f}{\beta}-1)[\hat{J}_{3},\hat{J}_{2}]+\hat{J}_{3}\hat{J}_{2} & (\hat{J}_{3})^{2} & -r'\hat{J}_{3}\cr
-r'\hat{J}_{1} & -r'\hat{J}_{2} & -r'\hat{J}_{3} & r'^{2}+\frac{M^{2}\lambda^{2}}{\beta^{2}}}
\nonumber
\end{eqnarray}
These results reduce to the previous ones when $M=0$ or, equivalently, $\beta=f$.

Now we introduce a parameter
\begin{equation}
\Delta\equiv \left(\frac{f}{\beta}-1\right)=\frac{M^{2}\lambda^{2}}{2\beta^{2}}
=\frac{1-\sqrt{1-\frac{2M^{2}\lambda^{2}}{f^{2}}}}{1+\sqrt{1-\frac{2M^{2}\lambda^{2}}{f^{2}}}}
\label{deformPara}
\end{equation}
to measure the mass deformation of our previous system. Using the 4 by 4 unitary matrix $U$ given in (\ref{unitary}), we find that the conjugate matrix $U^{\dag}{\cal N}_{B}U$ takes the form
\begin{eqnarray}
\pmatrix{\frac{1}{2}\hat{J}_{-}\hat{J}_{+}-2\Delta \hat{J}_{3}
& -\frac{1}{2}(\hat{J}_{-})^{2} & \frac{1}{\sqrt{2}}\hat{J}_{-}\hat{J}_{3}+\sqrt{2}\Delta \hat{J}_{-} & -\frac{r'}{\sqrt{2}}\hat{J}_{-}\cr
-\frac{1}{2}(\hat{J}_{+})^{2} & \frac{1}{2}\hat{J}_{+}\hat{J}_{-}+2\Delta \hat{J}_{3}
& -\frac{1}{\sqrt{2}}\hat{J}_{+}\hat{J}_{3}+\sqrt{2}\Delta \hat{J}_{+} & \frac{r'}{\sqrt{2}}\hat{J}_{+}\cr
\frac{1}{\sqrt{2}}\hat{J}_{3}\hat{J}_{+}+\sqrt{2}\Delta \hat{J}_{+} &
-\frac{1}{\sqrt{2}}\hat{J}_{3}\hat{J}_{-}+\sqrt{2}\Delta \hat{J}_{-} &
(\hat{J}_{3})^{2} & -r'\hat{J}_{3}\cr
-\frac{r'}{\sqrt{2}}\hat{J}_{+} & \frac{r'}{\sqrt{2}}\hat{J}_{-} & -r'\hat{J}_{3} & r'^{2}+2\Delta}
\nonumber
\end{eqnarray}
Thus, as discussed in Section~\ref{s2}, the eigenvectors of $\widetilde{\cal N}\equiv U^{\dag}{\cal N}_{B}U$ can be written as
\begin{eqnarray}
v_{\hat{j},\,\hat{m}}=\pmatrix{\alpha_{1}\,\Psi_{\hat{j},\;\hat{m}}\cr\alpha_{2}\Psi_{\hat{j},\hat{m}+2}\cr\alpha_{3}\Psi_{\hat{j},\hat{m}+1}\cr\alpha_{4}\Psi_{\hat{j},\hat{m}+1}},\quad \widetilde{\cal  N} v_{\hat{j},\,\hat{m}}=\Lambda v_{\hat{j},\,\hat{m}},
\label{deformEigenVect}
\end{eqnarray}
where $\Lambda$ is determined algebraically by
\begin{eqnarray}
\det\pmatrix{a_{1}b_{1}-2\Delta -\Lambda & a_{1}b_{2} & a_{1}b_{3}+c_{\hat{j}}(\hat{m})\Delta & a_{1}b_{4}\cr
a_{2}b_{1} & a_{2}b_{2}+2(\hat{m}+2)\Delta-\Lambda & a_{2}b_{3}+c_{\hat{j}}(\hat{m}+1)\Delta & a_{2}b_{4}\cr
a_{3}b_{1}+c_{\hat{j}}(\hat{m})\Delta & a_{3}b_{2}+c_{\hat{j}}(\hat{m}+1)\Delta
& a_{3}b_{3}-\Lambda & a_{3}b_{4}\cr
b_{1} & b_{2} & b_{3} & b_{4}+2\Delta-\Lambda}=0,
\nonumber
\end{eqnarray}
with
\begin{equation}
c_{\hat{j}}(\hat{m})=\sqrt{2(\hat{j}-\hat{m})(\hat{j}+\hat{m}+1)}.
\end{equation}
The numerical coefficients $a_{1},\cdots, a_{3}$ and $b_{1},\cdots, b_{4}$ here are defined as in (\ref{algCoeff}).

Solving the above algebraic equation we find the following four eigenvalues:
\begin{equation}
\Lambda_{1}=2\Delta,\quad \Lambda_{2}=-2\hat{j}\Delta,\quad\Lambda_{3}= 2(\hat{j}+1)\Delta ,\quad\Lambda_{4}= \hat{j}(\hat{j}+1)+r'^{2}+2\Delta.
\label{deformEigenVal}
\end{equation}
Note that for $\Delta$ taking generic values, all these eigenvalues are nondegenerate, namely, there is only one eigenvector $(\alpha_{1},\alpha_{2},\alpha_{3},\alpha_{4})$ associated to a given $\Lambda$. Remarkably, these eigenvalues are all independent of $\hat{m}$, so the eigenstates (\ref{deformEigenVect}) corresponding to each value in (\ref{deformEigenVal}) are $(2\hat{j}+1)$-fold degenerate.

The bosonic mass matrix ${\cal M}_{B}^{2}$ for the modes $A_{1},A_{2},A_{3},A_{4}$ thus has the spectrum
\begin{equation}
\frac{\lambda^{2}}{\beta^{2}}\tilde{\omega}_{B}^{2}=r'^{2}+\hat{j}(\hat{j}+1),\quad r'^{2}+(\hat{j}+2\Delta)(\hat{j}+1),\quad r'^{2}+\hat{j}(\hat{j}-2\Delta+1),\quad 0,
\label{DspecB1}
\end{equation}
of which each has the degeneracy $2\hat{j}+1$. In addition, the spectrum for the remaining bosonic modes $A_{5},A_{6},A_{7},A_{8},A_{9}$ can be read off directly from (\ref{energyB1}):
\begin{equation}
\frac{\lambda^{2}}{\beta^{2}}\omega_{B}^{2}=r'^{2}+\hat{j}(\hat{j}+1),
\label{DspecB2}
\end{equation}
and this eigenvalue is of $5(2\hat{j}+1)$-fold degeneracy. Finally, since adding the mass deformation terms to the action will not change the fermionic part of the energy much (only the classical values of $X^{a}$ are rescaled from $fY^{a}$ to $\beta Y^{a}$), the mass spectrum for the fermionic modes is simply given by
\begin{equation}
\frac{\lambda^{2}}{\beta^{2}}\omega_{F}^{2}=r'^{2}+(\hat{j}+1)^{2},\quad
r'^{2}+\hat{j}^{2},
\label{DspecF}
\end{equation}
of which each has $4(2\hat{j}+1)$-fold degeneracy.

Now we are ready to compute the zero-point energy of the deformed system. Using the frequencies $\omega_{B}$, $\tilde{\omega}_{B}$ and $\omega_{F}$ given in (\ref{DspecB1})-(\ref{DspecF}) (together with their degeneracies), we find:
\begin{eqnarray}
V(r)&=&\frac{\beta}{\lambda}\sum_{\hat{j}=0}^{N-1}(2\hat{j}+1)\left[6\sqrt{\hat{j}(\hat{j}+1)+r'^{2}}+\sqrt{(\hat{j}+2\Delta)(\hat{j}+1)+r'^{2}}\right.\nonumber\\
&+&\left.\sqrt{\hat{j}(\hat{j}-2\Delta+1)+r'^{2}}
-4\sqrt{(\hat{j}+1)^{2}+r'^{2}}-4\sqrt{\hat{j}^{2}+r'^{2}}\right]\nonumber\\
&=& \frac{\beta}{\lambda}\sum_{\hat{j}=0}^{N-1}(2\hat{j}+1)\frac{\Delta -2}{r'}\;=\;\frac{(\Delta-2)N^{2}\beta^{2}}{\lambda r}+{\cal O}(r^{-2}).
\label{ZeroEnergy2}
\end{eqnarray}
So once again we get the $1/r$-law of interactions, provided $\Delta\neq 2$.

In the above quantum mechanical system one has effectively $n_{b}=8$ bosons and $n_{f}=8$ fermions, and the coefficients $a_{i}({\bf k})$, $b_{i}({\bf k})$, $c_{i}({\bf k})$, $d_{i}({\bf k})$ in (\ref{generic}) can be identified with
\begin{equation}
a_{i}({\bf k})=c_{i}({\bf k})=\frac{1}{\lambda}\,(2\hat{j}+1),\quad (1\leq i\leq n_{b}=8)
\label{ac-coeff}
\end{equation}
\begin{equation}
b_{1}({\bf k})=\beta^{2}\,(\hat{j}+2\Delta)(\hat{j}+1),\quad b_{2}({\bf k})=\beta^{2}\,\hat{j}(\hat{j}-2\Delta+1),
\label{b-coeff1}
\end{equation}
\begin{equation}
b_{i}({\bf k})=\beta^{2}\,\hat{j}(\hat{j}+1),\quad\quad\quad \hbox{for $i=3,4,5,6,7,8$},
\label{b-coeff2}
\end{equation}
\begin{equation}
d_{i}({\bf k})=\beta^{2}\times\left\{
                \begin{array}{ll}
                   (\hat{j}+1)^{2},\quad  & \hbox{for $i=1,3,5,7$;}\\
                    \hat{j}^{2},    &   \hbox{for $i=2,4,6,8$.}
                \end{array}                  
\right.
\label{d-coeff}
\end{equation}
Note that the quantity $b_{i}({\bf k})$ is different from $d_{i}({\bf k})$ for $i\geq 3$; hence the first identity in (\ref{ac-coeff}) implies that the potential should behave as $1/r$. Indeed, as discussed before, the existence of dielectric effects requires $0\leq M^{2}\lambda^{2}/f^{2}\leq 1/2$, so the parameter $\Delta$ defined by (\ref{deformPara}) must range from 0 to 1. Consequently the factor $(\Delta-2)$ associated to the $1/r$ term in the zero-point energy (\ref{ZeroEnergy2}) never vanishes.

The above results are valid for any mass deformation parameter $\Delta\in[0,1]$. Given a generic value of $\Delta$, we see from (\ref{b-coeff1})-(\ref{d-coeff}) that for every $i=1,\cdots,8$, the quantity $b_{i}({\bf k})$ cannot be the same as $d_{i}({\bf k})$. Thus, introducing a generic mass parameter to the action of dielectric branes does not automatically make the system supersymmetric. This can be seen even at the classical level,  at which the ground state energy (\ref{ClassicEngergy}) in general has a nonvanishing value. However, when the mass parameter takes its critical value (\ref{SUSY}), the classical energy reaches zero at its minimum. In that case,  one simply deduces from (\ref{SUSY}) and (\ref{deformPara}) that 
\begin{equation}
\Delta=\frac{1-\sqrt{1-2\cdot\frac{4}{9}}}{1+\sqrt{1-2\cdot\frac{4}{9}}}=\frac{1}{2}.
\end{equation}
Putting this special value of $\Delta$ into (\ref{b-coeff1}), we find that $b_{1}({\bf k})$ and $b_{2}({\bf k})$ now become identical to $d_{1}({\bf k})$ and $d_{2}({\bf k})$, respectively. Consequently, all the quantized modes associated with the first two pairs of bosons and fermions cancel out in the zero-point energy. One possible implication of this fact is that SUSY is partially restored at the one-loop level.
\section{Dielectric 2-brane/D0-brane Interactions} \label{s4}
In this section we consider the long-range interaction between a dielectric 2-brane and a single D0-brane. The computations are nearly the same as those presented in Section~\ref{s2}, so our discussion here will be quite brief.

The classical configuration of the Dielectric 2-brane/D0-brane system can be described by the following $(N+1)\times(N+1)$ matrices:
\begin{equation}
X^{i} = \pmatrix{  fJ_{i}   & 0 \cr
                  0  & 0}, \quad 
X^{4} = \pmatrix{  0 & 0 \cr
                 0 & r}, \quad
		 X^{\mu}=0,\quad \theta=0
\label{0-classical}
\end{equation}
where the angular momentum $J_{i}$ belongs to the $N$-dimensional irreducible representation of $SU(2)$. For simplicity we make the rescaling $X^{a}=fY^{a}$, $\theta=T_{0}^{-1/2}\theta'$ and $r=fr'$. Quantum mechanically, we will add off-diagonal modes to (\ref{0-classical}) and investigate their excitations. Thus, let us introduce the $N\times 1$ column matrices $A_{a}$ representing quantum interactions between the dielectric brane and the D0 brane, with $ A^{\dag}_{i}$ being the corresponding $1\times N$ row matrices. We then insert the quantities
\begin{equation}
Y^{i} = \pmatrix{  J_{i}   & A_{i} \cr
                 A^{\dag}_{i}  & 0}, \quad 
Y^{4} = \pmatrix{  0 & A_{4} \cr
                 A^{\dag}_{4} & r'}, \quad
Y^\mu = \pmatrix{
                   0       & A_{\mu} \cr
                   A^{\dag}_{\mu} & 0}, \quad
\theta' = \pmatrix{
                   0 & \psi \cr
                      0 & 0}
\end{equation}
into the action of dielectric 2-branes. Keeping only quadratic terms we arrive at the following field theory expressions:
\begin{eqnarray}
V_{B}&=&\frac{T_{0}f^{4}}{\lambda^{2}}\int\left[A^{\dag}_{\mu}(J^{2}+r'^{2})A_{\mu}+{\bf A}^{\dag}{\cal M}_{B}^{2}{\bf A}\right],\nonumber\\
V_{F}&=&\frac{f}{\lambda}\int \psi^{T}{\cal M}_{F}\,\psi,
\end{eqnarray}
where the bosonic and fermionic mass matrices ${\cal M}_{B}$, ${\cal M}_{F}$ are defined as in (\ref{MMB}) and (\ref{MMF}), respectively, with the replacement of $\hat{J}_{i}$ by $J_{i}$. The mass spectra can be determined by solving the eigenvalue problems for these operator-valued matrices, and the result reads
\begin{eqnarray}
\omega_{B}^{2}&=&\frac{f^{2}}{\lambda^{2}}\times\left\{
                \begin{array}{ll}
                  j(j+1)+r'^{2},\quad & \hbox{with $(3+5)(2j+1)$-fold degeneracy,}\\
                    0, & \hbox{with $(2j+1)$-fold degeneracy;}
                \end{array}                  
\right.
\label{0-specB}\nonumber\\
\omega_{F}^{2}&=&\frac{f^{2}}{\lambda^{2}}\times
\left\{
             \begin{array}{ll}
                 (j+1)^{2} + r'^{2} ,\quad & \hbox{with $4(2j+1)$-fold degeneracy,}  \\             
             j^{2} + r'^{2}, \quad & \hbox{with $4(2j+1)$-fold degeneracy.}
             \end{array}
          \right.
\label{0-specF}
\end{eqnarray}
Note that here $j=(N-1)/2$ is taken fixed, rather than running from $0$ to $N-1$. So finally we can write down our quantum mechanical result of the zero-point energy:
\begin{eqnarray}
V(r) &=& \frac{f}{\lambda}\,(2j+1)\left\{
8\sqrt{j(j+1) + r'^2} - 4\sqrt{(j+1)^2 + r'^2}- 4\sqrt{j^2 + r'^2} \right\}\nonumber\\
&=& \frac{2f}{\lambda}\,(2j+1) (-\frac{1}{r'})\;=\; -\frac{2Nf^{2}}{\lambda\, r}+{\cal O}(\frac{1}{r^{2}}).
\label{0-ZeroEnergy}
\end{eqnarray}
\section{Conclusions} \label{s5}
In this paper we have computed the long-range interactions between two dielectric 2-branes of the same radius, as well as the zero-point energy of the D0-brane/dielectric 2-brane system. We found that the potentials behave as $1/r$ at large distance. We also investigated the effect of mass terms added to the action of dielectric branes. When the mass parameter takes a special value such that the classical energy vanishes at its minimum, we observed some interesting cancellations between bosons and fermions, which suggests that the deformed system may have a kind of supersymmetry. It would be nice to have a direct understanding of this phenomenon beyond the 1-loop calculations.

The computations presented here can be easily extended to the case of two dielectric branes with different radii. For example, if the first D0-D2 bound state consists of $N_{1}$ D0-branes and the second contains $N_{2}$ D0-branes, then the interaction between them is simply given by
\begin{equation}
V(r)=-\frac{2N_{1}N_{2}f^{2}}{\lambda\,r}+{\cal O}(\frac{1}{r^{2}}).
\end{equation}
This formula clearly interpolates between our previous results (\ref{ZeroEnergy1}) and (\ref{0-ZeroEnergy}).

Also, it would be possible to extend our computations to some more complicated dielectric branes, such as the fuzzy coset configurations \cite{TV}. A naive guess is that the sum in (\ref{generic}) should now be extended over the set of irreducible representations $\bf k$ arising in the decomposition of the tensor product $j_{1}\otimes j_{2}$, where $j_{1}$, $j_{2}$ label the highest weights associated with the first and the second coset spaces. Moreover, the coefficients $a_{i}({\bf k})$, $b_{i}({\bf k})$, $c_{i}({\bf k})$, $d_{i}({\bf k})$ in (\ref{generic}) should correspond to some simple group-theoretic objects. 

Another possible extension is to study interactions between two dielectric branes in the magnetic analog \cite{DTV} of Myers' theory. In particular, it would be interesting to consider the case when one fuzzy direction is wrapped around the M-circle, and find out some physical interpretations of the interactions in terms of M-theory.
\section *{Acknowledgement}
We would like to thank B.~Y.~Hou, M.~Li and M.~Yu for discussions.

\end{document}